\documentclass[12pt]{article}
\usepackage{amssymb}
\usepackage{amsmath}
\usepackage{amsthm}
\usepackage{graphicx}
\newtheorem{theorem}{Theorem}
\newtheorem{lemma}{Lemma}
\newtheorem{definition}{Definition}
\newtheorem{corollary}{Corollary}
\begin{document}
\title{Relative Unitary Implementability of Perturbed Quantum Field Dynamics on de-Sitter Space}
\author{Gary K. Poon\\{garypoon@buffalo.edu}\\Department of Physics, SUNY at Buffalo, Amherst, NY, 14260}
\maketitle
\begin{abstract}
In this article, we study the quantum dynamics of a Klein-Gordon field on de-Sitter space.  We prove time evolution is not unitarily implementable.  We also consider a Klein-Gordon field perturbed by a local potential $V$.  In this case we prove that the deviation from the $V=0$ dynamics is unitarily implementable,
\end{abstract}
\section{Introduction}
Our expanding universe manifold is roughly described as a de-Sitter space.  De-Sitter space is also worthy of attention because it is highly symmetric.  This facilitates the study of quantum field phenonmenon.  We study in particular the scalar field $\phi$ which is a solution of the Klein-Gordon equation,\\$(-\Box_g + m^2)\phi=0$.  This has been studied by many authors\cite{Chernikov, Nachtmann, Allen, Schomblond, Bros, Kay}.

We give a mathematical construction of $\phi$ which follows the treatment of Schomblond and Spindel\cite{Schomblond}.  The construction is based on the symmetric Euclidean vaccuum which has O(1,4) invariance.  We demonstrate this invariance as well as locality properties of the field.

We consider the issue of whether time evolution is unitarily implementable.  If $\phi(t,\vec{x})$ is the field at time $t$ and position, $\vec{x}$, is there a unitary operator $\mathcal{U}_0(t,t_0)$ so that $\phi(t,\vec{x}) = \mathcal{U}_0(t,t_0)\phi(t_0,\vec{x})\mathcal{U}_0^{-1}(t,t_0)$? We prove that no such $\mathcal{U}_0$ exists, a widely expected result.  This is an obstacle to giving a particle interpretation to the theory.

We consider also the Klein-Gordon field in the presence of external potential, satisfying $(-\Box_g + m^2 + V)\phi=0$, V a local scalar potential.  The classical time evolution operator $U_V(t,t_0)$ will be studied by an expansion in a Dyson series.  This makes possible the definition of a perturbed quantum field $\phi_V(t,\vec{x})$.  We show that the relative time evolution is unitarily implementable in the sense that there exists an unitary operator, $\mathcal{U}_V(t)$ such that $\phi_V(t,\vec{x})=\mathcal{U}_V(t)\phi(t,\vec{x})\mathcal{U}_V^{-1}(t)$. This is a new result.

The theory developed here may possibly be extended to treat other perturbations such as a local vector potential or a local change in the metric.  The hope is that unitary implementability will be a good technical tool for such investigations.
\section{Klein-Gordon Equation in de-Sitter Space}
We start with the de-Sitter metric in $(0,\infty)\times\mathbb{R}^3$ as fallows
\begin{equation*}
g_{\mu\nu} = \frac{R^2}{t^2}
	\begin{pmatrix}
	-1 & 0 &0 &0\\
	0 & 1 & 0 & 0\\
	0 & 0 & 1 & 0\\
	0 & 0 & 0 & 1
	\end{pmatrix}
, \quad
g^{\mu\nu} = \frac{t^2}{R^2}
	\begin{pmatrix}
	-1 & 0 &0 &0\\
	0 & 1 & 0 & 0\\
	0 & 0 & 1 & 0\\
	0 & 0 & 0 & 1\\
	\end{pmatrix}
\end{equation*}
where $R$ equals the inverse of the Hubble constant, or the radius of the universe.  De-Sitter space satisfies two cosmological principles, namely, homogeneity and isotropy.  It is also an expanding universe that satisfies Einstein's field equation with the cosmological constant, $\Lambda =H^2$ in natural units, where $H=R^{-1}$ is the Hubble constant:
\begin{equation*}
R_{\mu\nu}-\frac{1}{2}Rg_{\mu\nu} + \Lambda g_{\mu\nu}= 0
\end{equation*}
The right hand side of this equation is zero which corresponds to no matter existing in de-Sitter space.  The extra term $\Lambda g_{\mu\nu}$ comes from the hyporthesis of dark energy to explain the expansion rate of the universe.  The $t$ coordinate is chosen to run backward in time so the big bang occurs at $t=\infty$.  The manifold is not complete since geodesics can run off at $t=\infty$.

Let $\phi = \phi(t, \vec{x})$ be the scalar field, defined on the manifold $(0, \infty) \times \mathbb{R}^3$, satisfying the Klein-Gordon equation,
\begin{equation}
(-\Box_g + m^2)\phi = 0
\end{equation}
where the subscript $g$ represents the de-Sitter metric. We have
\begin{equation*}
|det(g)|^{1/2} = \left(\frac{R}{t}\right)^4
\end{equation*}
The D'Alembertian for the de-Sitter metric is
\begin{flalign*}
\Box_g & = \sum_{\substack{\mu,\nu}}|det(g)|^{-1/2}\frac{\partial}{\partial x_\mu}\left(|det(g)|^{1/2}g^{\mu\nu}\frac{\partial}{\partial x_\nu}\right) \\
       & = t^4\sum_{\substack{\mu,\nu}}\frac{\partial}{\partial x_\mu}\left(t^{-4} g^{\mu\nu}\frac{\partial}{\partial x_\nu}\right) \\
       & = t^4\frac{\partial}{\partial t}\left(-\frac{1}{R^2t^2}\frac{\partial}{\partial t}\right) + \frac{t^2}{R^2}\triangle \\
       & = -\frac{t^2}{R^2}\left(\frac{\partial^2}{\partial t^2} - \triangle\right) + \frac{2t}{R^2}\frac{\partial}{\partial t} 
\end{flalign*}
With this D'Alembertian, Klein-Gordon equation becomes
\begin{equation}
\left[\frac{t^2}{R^2}\left(\frac{\partial^2}{\partial t^2} - \triangle\right) - \frac{2t}{R^2}\frac{\partial}{\partial t} + m^2\right]\phi(t, \vec{x}) = 0
\end{equation}
We look for solutions of the form $\phi(t, \vec{x}) = e^{i\vec{k}\cdot\vec{x}}v(t,k)$ where $k=|\vec{k}|$.  Then (2) becomes
\begin{equation}
\left[t^2\frac{d^2}{dt^2} - 2t\frac{d}{dt} + (kt)^2 + (mR)^2\right]v(t,k) = 0
\end{equation}
Try $v(t,k) = t^{3/2}Z(kt)$. We have
\begin{flalign*}
v'(t,k) & = \frac{3}{2}t^{1/2}Z(kt) + t^{3/2}kZ'(kt)\\
v''(t,k) & = \frac{3}{4}t^{-1/2}Z(kt) + 3t^{1/2}kZ'(kt) + t^{3/2}k^2Z''(kt)
\end{flalign*}
Then (3) becomes
\begin{equation}
\left[x^2\frac{d^2}{dx^2} + x\frac{d}{dx} + x^2 + (mR)^2 - \frac{9}{4}\right]Z(x) = 0
\end{equation}
where $x=kt$.  This is Bessel's equation with the solution,
\begin{flalign*}
Z(kt) & = J_\nu(kt) \pm iN_\nu(kt)\\
& = H_\nu^{(1)}(kt) \text{ or } H_\nu^{(2)}(kt)
\end{flalign*}
where $\nu = \sqrt{\frac{9}{4} - (mR)^2}$.  We assume $mR>3/2$, so $\nu$ is purely imaginary as supported by empirical values \cite{Gazeau}, $mR=3.33\times 10^{37}$ for electron and $5\times 10^{43}$ for W$^\pm$ bosons.   $J_\nu(kt)$ and $N_\nu(kt)$ are Bessel's functions of the first kind and second kind, respectively and $H_\nu^{(1)}(kt) \text{, } H_\nu^{(2)}(kt)$ are the Hankel functions. The explicit expressions of Bessel and Hankel functions are given by Watson \cite{Watson}, 
\begin{equation}
\begin{split}
H_\nu^{(1)}(x) & = \frac{J_{-\nu}(x)-e^{-i\pi\nu}J_\nu(x)}{i\sin(\nu\pi)}\\
H_\nu^{(2)}(x) & = \frac{e^{i\pi\nu}J_\nu(x)-J_{-\nu}(x)}{i\sin(\nu\pi)}\\
J_\nu(x) & = \overset{\infty}{\sum_{\substack{n=0}}}\frac{(-1)^n}{n!(n+\nu)!}\left(\frac{x}{2}\right)^{\nu+2n}
\end{split}
\end{equation}
with convergence for all $x$.  One can rewrite the Bessel function as
\begin{equation*}
J_\nu(x)=x^{\nu}f_{\nu}(x)
\end{equation*}
where
\begin{equation*}
f_{\nu}(x)=2^{-\nu}\overset{\infty}{\sum_{\substack{n=0}}}\frac{(-1)^n}{n!(n+\nu)!}\left(\frac{x}{2}\right)^{2n}
\end{equation*}
is an entire analytic function.  Writing $\nu = i\mu,\mu = \sqrt{(mR)^2 - \frac{9}{4}}$, one notices $x^{\nu}$ is rapidly oscillating at $x=0$. From the explicit expression of Bessel function in (5), $\overline{J_\nu(x)} = J_{-\nu}(x)$.  Then we have
\begin{equation*}
\overline{H_\nu^{(1)}(kt)} = H_{-\nu}^{(2)}(kt)
\end{equation*}
and from definition in (5)
\begin{equation*}
H_\nu^{(1)}(kt) = e^{-i\pi\nu}H_{-\nu}^{(1)}(kt)
\end{equation*}
These two expressions yield
\begin{equation*}
\overline{H_\nu^{(1)}(kt)} = e^{\pi\mu}H_\nu^{(2)}(kt)
\end{equation*}
Then one has a pair of Hankel functions shifted by a constant such that
\begin{flalign*}
\mathcal{H}_\nu^{(1)}(kt)& = e^{-\mu\pi/2}H_\nu^{(1)}(kt)\\
\mathcal{H}_\nu^{(2)}(kt)& = e^{\mu\pi/2}H_\nu^{(2)}(kt)
\end{flalign*}
satisfying $\overline{\mathcal{H}_\nu^{(1)}(kt)}=\mathcal{H}_\nu^{(2)}(kt)$.  Now the solutions of (3) can be written as complex conjugate of each other as follow
\begin{equation}
\begin{split}
v(t, k) &= t^{3/2}\mathcal{H}_\nu^{(1)}(kt)\\
\bar{v}(t, k) &= t^{3/2}\mathcal{H}_\nu^{(2)}(kt)
\end{split}
\end{equation}
A general real solution has the form,
\begin{equation}
u(t, \vec{x})  = \frac{1}{(2\pi)^{3/2}}\int e^{i\vec{k}\cdot\vec{x}}[v(t,k)\psi(\vec{k}) + \bar{v}(t,k)\overline{\psi(-\vec{k})}]\mathrm{d}\vec{k}
\end{equation}

We now want to find an explicit expression for $\psi(\vec{k})$ so that the solution (7) gives Cauchy data, 
\begin{equation}
(f(\vec{x}), h(\vec{x})) = \left(u(t_0,\vec{x}), \partial u/\partial n(t_0, \vec{x})\right)
\end{equation}
 where $n$ is the forward unit normal vector on the Cauchy surface at $t=t_0$.  The normal derivative is written as $\partial u/\partial n = n^{\mu}\partial u/\partial x^{\mu}$.  The unit normal vector has the form, $n^\mu=(n^0,0,0,0)$ and satisfies
\begin{equation*}
1  = -g_{t_0}(n,n) = -g_{\mu\nu}(t_0,\vec{x})n^{\mu}n^{\nu} = \left(\frac{R}{t_0}\right)^2(n^0)^2
\end{equation*} 
so $n^0 =t_0/R$ and
\begin{equation*}
\frac{\partial}{\partial n} = \frac{t_0}{R}\frac{\partial}{\partial t}\bigg|_{t_0}
\end{equation*}  
Now the Cauchy data become $(f, h)= (u(t_0,\cdot), t_0/R\partial u/\partial t(t_0,\cdot))$.  For a general $t$, the Cauchy data become $(f_t, h_t) = (u(t,\cdot), t/R\partial u/\partial t(t,\cdot))$.  From (7), we have
\begin{flalign}
\tilde{f}_t(\vec{k}) & = v(t,k)\psi(\vec{k}) + \bar{v}(t,k)\overline{\psi(-\vec{k})}\\
R\tilde{h}_t(\vec{k}) & = t[v'(t,k)\psi(\vec{k}) + \bar{v}'(t,k)\overline{\psi(-\vec{k})}]
\end{flalign}
Solving the two equations (9) and (10) simultaneously, we get
\begin{equation}
\psi(\vec{k}) = \frac{1}{tW(t,k)}\text{det } \begin{pmatrix} \tilde{f}_t(\vec{k}) & \bar{v}(t,k)\\ R\tilde{h}_t(\vec{k}) & t\bar{v}'(t,k) \end{pmatrix}
\end{equation}
where the Wronskian is
\begin{equation*}
 W(t,k) = det \begin{pmatrix} v(t,k) & \bar{v}(t,k)\\ v'(t,k) & \bar{v}'(t,k) \end{pmatrix}
\end{equation*}
According to (6)
\begin{equation}
\begin{split}
W(t, k) & = t^3 kW[H_\nu^{(1)}(kt), H_\nu^{(2)}(kt)]\\
        & = -\frac{4i}{\pi} t^2  
\end{split}
\end{equation}
Since $W[H^{(1)}(x), H^{(2)}(x)] = -\frac{4i}{\pi x}$.\cite{Watson}  Note $W(t,k)$ does not depend on $k$.  We will see shortly that $v(t,k)$ is bounded with derivatives in $t$ polynomially bounded in $k$.  Then if $f, h \in \mathcal{S}(\mathbb{R}^3)$, the Schwartz space of smooth rapidly decreasing function, then $\psi$ is rapidly decreasing and (7) does give a true solution with these data.
\section{Time Evolution Operator}
In this section we develop estimates on the time evolution operator.  To begin we have an estimate on $v(t,k)$ in (6).
\begin{lemma}
For $t\geq1$,
\begin{flalign*}
|v(t, k)|& \leq C t^{3/2} \omega^{-1/2} (k)\\
|v'(t, k)| & \leq Ct \omega^{1/2} (k)\\
|v''(t, k)| & \leq Ct \omega^{3/2} (k)
\end{flalign*}
where $\omega(k)=\sqrt{1+k^2}$ and $C$ is a constant different from expression to expression.
\end{lemma}
\begin{proof}
First consider
\begin{equation*}
|v(t,k)| = e^{-\mu\pi/2}t^{3/2} |H_\nu^{(1)}(kt)|
\end{equation*}
For $|x| \leq 1$, the Hankel function $H_\nu^{(1)}(x)$ is written in terms of the Bessel function $J_\nu(x)$, as seen in (5), which is a convergent series with a radius of convergence $\infty$. Hence $J_\nu(x)$ and $J_{-\nu}(x)$ are bounded for $|x| \leq 1$ and $H_\nu^{(1)}(x)$, $H_\nu^{(2)}(x)$ are bounded functions as well.  For $|x| > 1$, one has the asymptotic expansion of Hankel function \cite{Watson},
\begin{equation*}
H_\nu^{(1)}(x)  = \sqrt{\frac{2}{\pi x}}e^{i[x-(\nu + \frac{1}{2})\frac{\pi}{2}]}\left[\overset{k}{\sum_{\substack{n=0}}}\frac{(-1)^n(\nu, n)}{(2ix)^n} + \mathcal{O}\left(\frac{1}{x^{k+1}}\right)\right]
\end{equation*}
where
\begin{equation*}
(\nu, n)=\frac{\Gamma(\nu+n+\frac{1}{2})}{n!\Gamma(\nu-n+\frac{1}{2})}
\end{equation*}
Then
\begin{equation*}
|H_\nu^{(1)}(x)| = \sqrt{\frac{2}{\pi x}}e^{\frac{\pi}{2}\mu}[1+\mathcal{O}(|x|^{-1})]
\end{equation*}
Hence, the Hankel function is a bounded function such that $|H_\nu^{(1)}(x)| \leq C|x|^{-1/2}$ where $C>0$.  Then the Hankel function is bounded as follow
\begin{flalign*}
|H_\nu^{(1)}(x)| &\leq C 
\begin{cases}1 & \quad \text{for } |x| \leq 1\\ |x|^{-1/2} & \quad \text{for } |x| > 1 \end{cases}\\
&\leq C \omega^{-1/2}(x)\\
\end{flalign*}
So 
\begin{equation*}
|H_\nu^{(1)}(kt)|\leq C\omega^{-1/2}(k)
\end{equation*}
where I used the fact $\omega(kt)\geq\omega(k)$ for $t\geq1$.  Then we have the bounded value for $|v(t,k)|\leq Ct^{3/2}\omega^{-1/2}(k)$.

Second, 
\begin{equation*}
v'(t,k) = e^{-\mu\pi/2}\left(\frac{3}{2}t^{1/2}H_\nu^{(1)} (kt) + t^{3/2}kH_\nu^{(1)'}(kt)\right)
\end{equation*}
where $H_\nu^{(1)}(kt)$ has been shown bounded for all values of $kt$.  What is left is to show the boundedness of
\begin{equation*}
H_\nu^{(1)'}(x) = \frac{J_{-\nu}'(x)-e^{-i\pi\nu}J_\nu'(x)}{i\sin(\nu\pi)}
\end{equation*}
Using the rewritten Bessel function following from expressions (5),
\begin{equation*}
J_\nu'(x)= \nu x^{\nu-1}f_{\nu}(x) + x^{\nu}f_{\nu}'(x)
\end{equation*}
Then for $|x|\leq1$
\begin{equation*}
|J_\nu'(x)|=\mathcal{O}(|x|^{-1})
\end{equation*}
so that $|H_\nu^{(1)'}(x)|\leq C|x|^{-1}$ for $|x|\leq 1$.  For $|x| > 1$, one uses the identity, \cite{Watson}
\begin{equation}
H_\nu^{(1)'}(x) = \frac{1}{2}[H_{\nu-1}^{(1)}(x) - H_{\nu+1}^{(1)}(x)]
\end{equation}
Then
\begin{equation*}
|H_\nu^{(1)'}(x)| \leq \frac{1}{2}[|H_{\nu-1}^{(1)}(x)|+|H_{\nu+1}^{(1)}(x)|]
\end{equation*}
As shown previously, the asymptotic expansions of Hankel functions, $H_{\nu-1}^{(1)}(x)$ and $H_{\nu+1}^{(1)}(x)$ are bounded by $C|x|^{-1/2}$. Hence, $|H_\nu^{(1)'}(x)|\leq C|x|^{-1/2}$ where $C>0$.  One then has a bounded $|H_\nu^{(1)'}(x)|$
\begin{equation*} 
|H_\nu^{(1)'}(x)| \leq C|x|^{-1}
\begin{cases}1 & \quad \text{for } |x| \leq 1\\ |x|^{1/2} & \quad \text{for } |x| > 1 \end{cases}
\end{equation*}
With the bounded value of $|H_\nu^{(1)}(x)|$ found previously, one has
\begin{flalign*} 
|v'(t,k)| &\leq Ct^{1/2}
\begin{cases}1 & \quad \text{for } kt \leq 1\\ (kt)^{1/2} & \quad \text{for } kt > 1 \end{cases}\\
&\leq Ct\omega^{1/2}(k)\\
\end{flalign*}

Third, 
\begin{equation*}
v''(t,k) = e^{-\mu\pi/2}[\frac{3}{4}t^{-1/2}H_\nu^{(1)}(kt) + 3t^{1/2}kH_\nu^{(1)'}(kt) + t^{3/2}k^2H_\nu^{(1)''} (kt)]
\end{equation*}
For $|x|>1$, one uses the bound on $H_\nu^{(1)}(x)$ and the identity (13) as follows
\begin{flalign*}
|H_\nu^{(1)''}(x)| & \leq \frac{1}{2}[|H_{\nu-1}^{(1)'}(x)| + |H_{\nu+1}^{(1)'}(x)|]\\
& \leq \frac{1}{4}[|H_{\nu-2}^{(1)}(x)|+2|H_{\nu}^{(1)}(x)|+|H_{\nu+2}^{(1)}(x)|]\\
& \leq C|x|^{-1/2}
\end{flalign*}
And one uses the definitions of Hankel function and Bessel function from (5) for $|x|\leq1$,
\begin{equation*}
H_\nu^{(1)''}(x) = \frac{J_{-\nu}''(x)-e^{-i\pi\nu}J_\nu''(x)}{i\sin(\nu\pi)}
\end{equation*}
and
\begin{equation*}
J_{\nu}''(x) = \nu(\nu-1)x^{\nu-2}f_{\nu}(x) + 2\nu x^{\nu-1}f_{\nu}'(x) +x^{\nu}f_{\nu}''(x)
\end{equation*}
Then one has
\begin{equation*}
|J_{\nu}''(x)| = \mathcal{O}(|x|^{-2})
\end{equation*}
so that
\begin{equation*}
|H_{\nu}^{(1)''}(x)| \leq C|x|^{-2}
\end{equation*}As a result,
\begin{equation*} 
|H_\nu^{(1)''}(x)| \leq C|x|^{-2}
\begin{cases}1 & \quad \text{for } |x| \leq 1\\ |x|^{3/2} & \quad \text{for } |x| > 1 \end{cases}
\end{equation*}
With the bounded values of $|H_\nu^{(1)}(x)|$ and $|H_\nu^{(1)'}(x)|$ found previously, one has
\begin{flalign*} 
|v''(t,k)| &\leq Ct^{-1/2}
\begin{cases}1 & \quad \text{for } kt \leq 1\\ (kt)^{3/2} & \quad \text{for } kt > 1 \end{cases}\\
&\leq Ct\omega^{3/2}(k)
\end{flalign*}
\end{proof}
Now we can demonstrate the smoothness of our solutions.
\begin{lemma}
Let $f,h\in\mathcal{S}(\mathbb{R}^3)$.  Then the solution $u(t, \vec{x})$ with these data given by (7), (11) is $C^\infty$
\end{lemma}
\begin{proof}
Formally, the general spatial derivative $\partial_x^\alpha =\partial^{\alpha_1}_{x_1}\partial^{\alpha_2}_{x_2}\partial^{\alpha_3}_{x_3}$ on $u(t,\vec{x})$ is
\begin{equation*}
\partial_x^\alpha u(t, \vec{x})  = \frac{1}{(2\pi)^{3/2}}\int(ik)^\alpha e^{i\vec{k}\cdot\vec{x}}[v(t,k)\psi(\vec{k}) + \bar{v}(t,k)\overline{\psi(-\vec{k})}]\mathrm{d}\vec{k}
\end{equation*}
where $k^\alpha=k_1^{\alpha_1}k_2^{\alpha_2}k_3^{\alpha_3}$.  These are actual derivatives if all integrals are absolutely convergent.  Since $k^\alpha$ and $v(t,k)$ from lemma 1 are polynomially bounded in $k$ and $\psi(\vec{k})$ is Schwartz function(rapidly decreasing), therefore integrands are all absolutely convergent.

Formally,
\begin{equation*}
\partial_t^nu(t, \vec{x})  = \frac{1}{(2\pi)^{3/2}}\int e^{i\vec{k}\cdot\vec{x}}\left[\frac{\partial^n}{\partial t^n}v(t,k)\psi(\vec{k}) + \frac{\partial^n}{\partial t^n}\bar{v}(t,k)\overline{\psi(-\vec{k})}\right]\mathrm{d}\vec{k}
\end{equation*}
We need to show $\frac{\partial^n}{\partial t^n}v(t,k)$ to be polynomially bounded in $k$.  It is enough to show $\frac{\partial^n}{\partial t^n}\mathcal{H}_\nu(kt)$ is polynomially bounded in $k$ and we have
\begin{equation*}
\frac{\partial^n}{\partial t^n}\mathcal{H}_\nu(kt) = k^n\frac{\mathrm{d}^n}{\mathrm{d}x^n}\mathcal{H}_\nu(x)\bigg|_{x=kt}
\end{equation*}
But $\frac{\mathrm{d}^n}{\mathrm{d}x^n}\mathcal{H}_\nu(x)$ is a sum of $\mathcal{H}_{\nu\pm j}(x)$ for $0\leq j\leq n$, and $\mathcal{H}_{\nu\pm j}(x)$ satisfies $\mathcal{H}_{\nu\pm j}(x)\leq C\omega^{-1/2}(x)$ just as for $j=0$.  Hence the result.
\end{proof}
Let's look at the Sobolev spaces defined by 
\begin{flalign*}
H^{1/2}(\mathbb{R}^3)&=\{f\in L^2(\mathbb{R}^3) : \omega^{1/2}\tilde{f}\in L^2(\mathbb{R}^3)\}\\H^{-1/2}(\mathbb{R}^3)&=\{h\in L^2(\mathbb{R}^3) : \omega^{-1/2}\tilde{h}\in L^2(\mathbb{R}^3)\}
\end{flalign*}

We now define a real-linear map $K$ which takes real Cauchy data at $t$ to the coefficient $\psi$ given by expression (11).
\begin{equation}
[K_t(f,h)](\vec{k})  = \frac{1}{tW(t,k)}\text{det } \begin{pmatrix} \tilde{f}(\vec{k}) & \bar{v}(t,k)\\ R\tilde{h}(\vec{k}) & t\bar{v}'(t,k) \end{pmatrix}
\end{equation}
Also we define a real linear map $L_t$ which takes the coefficient $\psi$ to the Cauchy data.  With $u(t,\vec{x})$ given by (7),
\begin{equation}
\begin{split}
L_t\psi(\vec{k}) & =\left(u(t,\vec{x}), \frac{t}{R}\frac{\partial u}{\partial t}(t,\vec{x})\right)\\
& = \bigg(\frac{1}{(2\pi)^{3/2}}\int e^{i\vec{k}\cdot\vec{x}}[v(t,k)\psi(\vec{k}) + \bar{v}(t,k)\overline{\psi(-\vec{k})}]\mathrm{d}\vec{k},\\
& \quad \quad\frac{t}{R(2\pi)^{3/2}}\int e^{i\vec{k}\cdot\vec{x}}[v'(t,k)\psi(\vec{k}) + \bar{v}'(t,k)\overline{\psi(-\vec{k})}]\mathrm{d}\vec{k}\bigg)
\end{split}
\end{equation}
\begin{lemma}
$K_t$ is bounded from $H^{1/2}(\mathbb{R}^3) \oplus H^{-1/2}(\mathbb{R}^3)$ to $L^2(\mathbb{R}^3)$
\end{lemma}
\begin{proof}
We start with (14)
\begin{flalign*}
|K_t(f, h)(\vec{k})|& = \frac{\pi}{4t}\bigg|\text{det }\begin{pmatrix} \tilde{f} & \bar{v}(t,k)\\ R\tilde{h} & t\bar{v}'(t,k) \end{pmatrix}\bigg|\\
& \leq \frac{\pi}{4t}[|\tilde{f}(\vec{k})t\bar{v}'(t,k)| + |R\tilde{h}(\vec{k})\bar{v}(t,k)|]\\
& \leq C_t[|\tilde{f}(\vec{k})|\omega^{1/2}(k) + |\tilde{h}(\vec{k})|\omega^{-1/2}(k)]
\end{flalign*}
where I have used lemma 1 in the last line and $C_t=Ct$.  Now, using the inequality $(a+b)^2\leq 2(a^2 + b^2)$ for the absolute value square,
\begin{equation*}
|K_t(f, h)(\vec{k})|^2 \leq 2(C_t)^{2}[|\tilde{f}(\vec{k})|^2\omega(k) + |\tilde{h}(\vec{k})|^2\omega^{-1}(k)]
\end{equation*}
Integrated over $\vec{k}$,
\begin{equation*}
\|K_t(f, h)\|_2^2   \leq 2(C_t)^2(\|f\|^2_{H^{1/2}} + \|h\|^2_{H^{-1/2}})
\end{equation*}Or
\begin{equation*} 
\|K_t(f, h)\|_2 \leq \sqrt{2}C_t\|(f,h)\|_{H^{1/2}\oplus H^{-1/2}}
\end{equation*}
\end{proof}
\begin{lemma}
$L_t$ is bounded from $L^2(\mathbb{R}^3)$ to $H^{1/2}(\mathbb{R}^3) \oplus H^{-1/2}(\mathbb{R}^3)$
\end{lemma}
\begin{proof}
Let's take the Fourier transform of the expression of (15)
\begin{equation*}
\widetilde{L_t\psi}(\vec{k}) = ((\widetilde{L_t\psi})_1(\vec{k}),(\widetilde{L_t\psi})_2(\vec{k}))
\end{equation*}
where
\begin{flalign*}
(\widetilde{L_t\psi})_1(\vec{k}) & = v(t,k)\psi(\vec{k}) + \bar{v}(t,k)\bar{\psi}(-\vec{k})\\
(\widetilde{L_t\psi})_2(\vec{k}) & = \frac{t}{R}[v'(t,k)\psi(\vec{k}) + \bar{v}'(t,k)\bar{\psi}(-\vec{k})]
\end{flalign*}
Using lemma 1,
\begin{flalign*}
|(\widetilde{L_t\psi})_1(\vec{k})| & \leq Ct^{3/2}\omega^{-1/2}(|\psi(\vec{k})| + |\bar{\psi}(-\vec{k})|)\\
|(\widetilde{L_t\psi})_2(\vec{k})| & \leq \frac{Ct^2}{R}\omega^{1/2}(|\psi(\vec{k})| + |\bar{\psi}(-\vec{k})|)
\end{flalign*}
With these two inequalities, we have
\begin{flalign*}
\|L_t\psi\|^2_{H^{1/2}\oplus H^{-1/2}} & = \|((L_t\psi)_1,(L_t\psi)_2)\|^2_{H^{1/2}\times H^{-1/2}}\\
& = \|(L_t\psi)_1\|^2_{H^{1/2}} + \|(L_t\psi)_2\|^2_{H^{-1/2}}\\
& = \|\omega^{1/2}(\widetilde{L_t\psi})_1\|^2_2 + \|\omega^{-1/2}(\widetilde{L_t\psi})_2\|^2_2\\
& \leq 8(C_t)^2\|\psi\|^2_2
\end{flalign*}
where I have used the inequality $(a+b)^2\leq2(a^2+b^2)$ and $C_t=Ct^2$.  So,
\begin{equation*}
\|L_t\psi\|_{H^{1/2}\oplus H^{-1/2}} \leq \sqrt{8}C_t\|\psi\|_2
\end{equation*}
\end{proof}
Explicit calculation shows that
\begin{equation*}
K_tL_t \psi = \psi
\end{equation*}
And
\begin{equation*}
L_tK_t(f,h) = (f,h)
\end{equation*}
Then $K_t$ is a bijection and $L_t = K_t^{-1}$.  

We now come to the study of the time evolution operator, $U_0(t,s)$.  Given $f_s, h_s \in \mathcal{S}(\mathbb{R}^3)$, consider the solution of Klein-Gordon equation with this data as in (7).  Let $(f_t,h_t)$ be the data at some other time $t$ and $U_0(t,s)$ the linear map defined by
\begin{equation*}
U_0(t,s)(f_s, h_s) = (f_t, h_t)
\end{equation*}
then $\psi=K_s(f_s, h_s)$ and $(f_t, h_t)=K_t^{-1}\psi$.  So
\begin{equation}
U_0(t,s)=K_t^{-1}K_s
\end{equation}
But this operator is bounded on $H^{1/2}(\mathbb{R}^3) \times H^{-1/2}(\mathbb{R}^3)$ and $\mathcal{S}(\mathbb{R}^3) \times \mathcal{S}(\mathbb{R}^3)$ is dense in $H^{1/2}(\mathbb{R}^3) \times H^{-1/2}(\mathbb{R}^3)$ so this defines $U_0(t,s)$ as unique extension to $H^{1/2}(\mathbb{R}^3) \times H^{-1/2}(\mathbb{R}^3)$.  We take $U_0(t,s)$ as the basic dynamics.  We have the identities,
\begin{flalign*}
U(t,s)U(s,t) & = K_t^{-1}K_sK_s^{-1}K_t=I\\
U(t,u)U(u,s) & = K_t^{-1}K_uK_u^{-1}K_s\\
& = K_t^{-1}K_s\\
& = U(t,s)
\end{flalign*}
We now define the time evolution operator on the coefficient $\psi$ by
\begin{flalign*}
\hat{U}_0(t,1) &= K_1U_0(t,1)K_1^{-1}\\
& = K_1K_t^{-1}
\end{flalign*}
which is bounded on $L^2(\mathbb{R}^3)$ since $K_1$ and $K_t^{-1}$ are bounded operators proved by lemma 2 and 3, respectively. The usefulness of $\hat{U}_0(t,1)$ will become clearer as we proceed.  Now, let's find the explicit expression for $\hat{U}_0(t,1)$ for the sake of theorem 4 proved later on,
\begin{flalign*}
\hat{U}_0(t,1)\psi & = K_1 K_t^{-1}\psi\\
& = K_1 (f_t, h_t)\\
& = \frac{1}{W(1,k)}\text{det } \begin{pmatrix} \tilde{f}_t & \bar{v}(1,k)\\ R\tilde{h}_t & \bar{v}'(1,k) \end{pmatrix}
\end{flalign*}
where $\tilde{f}_t$ and $\tilde{h}_t$ are given by expressions (9) and (10).  Then
\begin{multline}
\hat{U}_0(t,1)\psi(\vec{k}) = \frac{1}{W(1,k)}\bigg[\text{det} \begin{pmatrix} v(t,k) & \bar{v}(1,k)\\ tv'(t,k) & \bar{v}'(1,k) \end{pmatrix}\psi(\vec{k})\\ + \text{det} \begin{pmatrix} \bar{v}(t,k) & \bar{v}(1,k)\\ t\bar{v}'(t,k) & \bar{v}'(1,k) \end{pmatrix}\overline{\psi(-\vec{k})}\bigg]
\end{multline}
\section{Time Evolution Operator with Potential}
In last section, the boundedness of the time evolution operator without the presence of potential is studied.  In this section, we will study the one in the presence of a scalar potential, $V(t,\vec{x})$.  First the Klein-Gordon equation needs to be rewritten as a first order linear differential equation in order to express the time evolution operator into Dyson series.  Let's begin with $\Phi = (\phi, \pi)$ where $\phi$ is a solution of the Klein-Gordon equation (1) and the conjugate field $\pi(\vec{x}, t)$ is defined as 
\begin{equation*}
\pi = \frac{\partial \phi}{\partial n} = \frac{t}{R}\frac{\partial \phi}{\partial t}
\end{equation*} 
so that
\begin{equation}
\frac{\partial \phi}{\partial t} = \frac{R}{t}\pi
\end{equation}
Taking the second derivative,
\begin{equation}
\frac{\partial^2 \phi}{\partial t^2} = \frac{R}{t}\frac{\partial \pi}{\partial t} - \frac{R}{t^2}\pi
\end{equation}
Substituting (18) and (19) into (2), we get
\begin{equation}
\frac{\partial \pi}{\partial t} = \frac{3}{t}\pi + \frac{t}{R}\triangle\phi - \frac{R}{t}m^2\phi
\end{equation}
Altogether then with $\Phi_t = (\phi_t, \pi_t)$ 
\begin{equation}
\frac{d\Phi_t}{dt} = -H_0(t)\Phi_t
\end{equation}
where the Hamiltonian
\begin{equation*}
H_0=\begin{pmatrix} 0 & -\frac{R}{t}\\ -\frac{t}{R}\triangle + \frac{R}{t}m^2 & -\frac{3}{t} \end{pmatrix}
\end{equation*}

In the presence of potential, $m^2$ replaced by $m^2 + V'(t,\vec{x})$, the Klein-Gordon equation (1) becomes $(-\Box_g + m^2 + V'(t,\vec{x}))\phi=0$ and with the results of (18) and (20), Klein-Gordon equation can be rewritten as
\begin{equation}
\frac{d\Phi_t}{dt} = -[H_0(t) + V(t)]\Phi_t
\end{equation}
where the perturbation
\begin{equation*}
V(t)=\begin{pmatrix} 0 & 0 \\ V(t,\vec{x}) & 0 \end{pmatrix}
\end{equation*}
and I have redefined $V(t,\vec{x})=\frac{R}{t}V'(t,\vec{x})$.  

Next we look for a time evolution operator $U_V(t, 1)$ in the presence of a potential.  If the field evolves as $\Phi_t=U_V(t, 1)\Phi$, (22) becomes
\begin{equation}
\begin{split}
\frac{dU_V}{dt}(t, 1)&=-[H_0(t) + V(t)]U_V(t, 1)\\
\text{and } U_V(1,1) &= I
\end{split}
\end{equation}
This equation can be formally solved if the potential $V$ is taken as a perturbation so that $U_V(t, 1)$ is expanded into a Dyson series,
\begin{equation}
U_V(t, 1) = U_0(t, 1) - \int_{1}^{t}U_0 (t,s)V(s)U_0(s,1)\mathrm{d}s +\cdots
\end{equation}
Again, we look at $U_V(t,1)$ transformed by $K=K_1$ that was defined in section 3, 
\begin{equation}
\begin{split}
\hat{U}_V(t,1) & = KU_V(t,1)K^{-1} \\
	       & = \hat{U}_0(t,1) - \int_{1}^{t}KU_0 (t,s)K^{-1}KV(s)K^{-1}KU_0(s,1)K^{-1}\mathrm{d}s +\cdots \\
	       & = \hat{U}_0(t,1) - \int_{1}^{t}\hat{U}_0(t,s)\hat{V}(s)\hat{U}_0(s,1)\mathrm{d}s +\cdots
\end{split}
\end{equation}
where $\hat{V}(s)=KV(s)K^{-1}$.  Including all higher order terms,
\begin{multline*}
\hat{U}_V(t,1) = \hat{U}_0(t,1)\\ \quad\quad\quad\quad\quad+\overset{\infty}{\sum_{\substack{n=1}}}(-1)^n\int_{1}^{t}\cdots\int_{1}^{s_{n-1}}\hat{U}_0(t,s_1)\hat{V}(s_1)\cdots\hat{V}(s_n)\hat{U}_0(s_n,1)\mathrm{d}s_n\cdots\mathrm{d}s_1
\end{multline*}

Next, we have a theorem for the convergence of the series expansion of $\hat{U}_V(t,1)$.
\begin{theorem}
For fixed $t$, we suppose $\|\tilde{V}(s,\cdot)\|_1 \leq C\text{ for } 1\leq s \leq t$.  Then the Dyson series for $\hat{U}_V(t,1)$ converges in $L^2(\mathbb{R}^3)$.
\end{theorem}
\begin{proof}
\begin{multline}
\|\hat{U}_V(t,1)\psi\| \leq \|\hat{U}_0(t,1)\psi\|\\ + \overset{\infty}{\sum_{\substack{n=1}}}\int_{1}^{t}\cdots\int_{1}^{s_{n-1}}\|\hat{U}_0(t,s_1)\|\|\hat{V}(s_1)\|\cdots\\\|\hat{V}(s_n)\|\|\hat{U}_0(s_n,1)\|\mathrm{d}s_n\cdots\mathrm{d}s_1\|\psi\| 
\end{multline}
The boundedness of $\hat{U}_V(t,1)$ is then analysed by each factor.  Let's look at the boundedness of the time evolution operators, $\hat{U}_0(s,s')$ as follows 
\begin{equation*}
\|\hat{U}_0(s,s')\psi\|_2 \leq \|K\|\|U_0(s,s')\|\|K^{-1}\|\|\psi\|_2
\end{equation*} where $\|K\|\leq C$ and $\|K^{-1}\|\leq C$ as proved by lemma 2 and 3. And in accordance with (16)
\begin{equation*}
\|U_0(s,s')\|\leq\|K_s^{-1}\|\|K_{s'}\|
\end{equation*}
where $\|K_{s'}\| \leq \sqrt{2}C_{s'}\leq\sqrt{2}C_{t}$ and $\|K_s^{-1}\|\leq\sqrt{8}C_s\leq\sqrt{8}C_t$ for $1 \leq s, s'\leq t$.  That implies $\|\hat{U}_0(s,s')\psi\|_2 \leq C_t\|\psi\|_2$, independent of $s$ and $s'$.  Next, the boundedness of $\hat{V}(s)$ follows from
\begin{equation}
\begin{split}
\hat{V}(s)\psi & = KV(s)K^{-1}\psi\\
& = K \begin{pmatrix} 0 & 0 \\ V(s,\cdot) & 0 \end{pmatrix}\left(\begin{array}{c}f_1\\h_1\end{array}\right)\\
& = K (0, V(s,\cdot)f_1)\\
& = -(2\pi)^{3/2}R\frac{\bar{v}(1)}{W(1)}\tilde{V}(s,\cdot)\ast\tilde{f}_1\end{split}
\end{equation}
where $f_1$ and $h_1$ are given by (9) and (10) respectively and the last step comes from the results of (14).  Taking the $L^2(\mathbb{R}^3)$ norm, (27) can be expressed as
\begin{equation}
\begin{split}
\|\hat{V}(s)\psi\|_2 & \leq (2\pi)^{3/2}R\|\frac{\bar{v}(1)}{W(1)}\|_{\infty}\|\tilde{V}(s, \cdot)\ast\tilde{f}_1\|_2\\
& \leq (2\pi)^{3/2}R\|\frac{\bar{v}(1)}{W(1)}\|_{\infty}\|\tilde{V}(s, \cdot)\|_1\|\tilde{f}_1\|_2
\end{split}
\end{equation}
where Young's theorem was used for the last step.  Now one can analyse the boundedness of $\hat{V}(s)$ by each factor, first $\|\tilde{V}(s,\cdot)\|_1$  is bounded by assumption.  Second $\|\tilde{f}_1\|_2 \leq [\|v(1,\cdot)\|_\infty + \|\bar{v}(1,\cdot)\|_\infty]\|\psi\|_2$ which comes from (9). As seen from lemma 1,
\begin{equation*}
|v(1, k)|=|\bar{v}(1,k)| \leq C\omega^{-1/2}(k) 
\end{equation*}
So, $\|v(1,\cdot)\|_\infty$ and $\|\bar{v}(1,\cdot)\|_\infty$ are finite and $\tilde{f}_1$ is bounded on $L^2(\mathbb{R}^3)$. Third $W^{-1}(1)\|\bar{v}(1,\cdot)\|_{\infty}$ is bounded since $\|\bar{v}(1,\cdot)\|_{\infty}$ is bounded and the Wronskian, $|W(1)| = \frac{4}{\pi}$, by (12). Hence, $\hat{V}(s)$ is bounded.  Since $\hat{U}_0(s_i,s_{i+1})$ is bounded by $C_t$ dependent on $t$ only and $\hat{V}(s_i)$ are bounded for $i=1,\cdots, n$, the integrand of the n-th term in (26) can be written as,\\ $\|\hat{U}_0(t,s_1)\|\|\hat{V}(s_1)\|\cdots\|\hat{V}(s_n)\|\|\hat{U}_0(s_n,1)\| \leq \Lambda_t^{n+1}$ where $\Lambda_t=\Lambda t^3$. Then, after integrating over the $s_i$, we have
\begin{flalign*}
\|\hat{U}_V(t,1)\psi\|_2 & \leq \Lambda_t\overset{\infty}{\sum_{\substack{n=0}}}\frac{\Lambda_t^n(t-1)^n}{n!}\|\psi\|\\
& = \Lambda_te^{\Lambda_t(t-1)}\|\psi\|_2
\end{flalign*}
\end{proof}
To complete the discussion of time evolution, we have the following corollary
\begin{corollary}
$U_V(t,1)$ defines a bounded operator on $H^{1/2}(\mathbb{R}^3) \times H^{-1/2}(\mathbb{R}^3)$ which solves (23).
\end{corollary}
\begin{proof}
\begin{equation*}
U_V(t,1) = K^{-1}\hat{U}_V(t,1)K
\end{equation*}
bounded from $H^{1/2}(\mathbb{R}^3) \times H^{-1/2}(\mathbb{R}^3)$ to $H^{1/2}(\mathbb{R}^3) \times H^{-1/2}(\mathbb{R}^3)$
\end{proof}
Before the end of this section, we introduce a sympletic form defined for a pair of test functions $F=(f,h)$ as,
\begin{equation*}
\sigma(F_1,F_2) = (f_1,h_2) - (h_1, f_2)
\end{equation*}
\begin{lemma}
$\sigma(F_1,F_2)$ is a well defined bounded bilnear form on $H^{1/2}(\mathbb{R}^3) \oplus H^{-1/2}(\mathbb{R}^3)$.
\end{lemma}
\begin{proof}
For the $3\times3$ spatial part of the de-Sitter metric, $g^{(3)}(t)=\left(\frac{R}{t}\right)^6$,
\begin{flalign*}
(f,h) & =\sqrt{\text{det}(g^{(3)}(t))}\int\overline{f(\vec{x})} h(\vec{x})\mathrm{d}\vec{x}\\
& = \left(\frac{R}{t}\right)^3\int \overline{\tilde{f}(\vec{k})}\tilde{h}(\vec{k})\mathrm{d}\vec{k}\\
&=\left(\frac{R}{t}\right)^3\int\overline{\omega^{1/2}(k)\tilde{f}(\vec{k})}\omega^{-1/2}(k)\tilde{h}(\vec{k})\mathrm{d}\vec{k}\\
& \leq\left(\frac{R}{t}\right)^3\|f\|_{H^{1/2}}\|h\|_{H^{-1/2}}\\
& < \infty
\end{flalign*}
For the boundedness we have
\begin{flalign*}
|\sigma(F_1,F_2)| & = |(f_1,h_2) - (h_1, f_2)|\\
& \leq \|f_1\|_{H^{1/2}}\|h_2\|_{H^{-1/2}}+\|h_1\|_{H^{-1/2}}\|f_2\|_{H^{1/2}}\\
&\leq\sqrt{\|f_1\|_{H^{1/2}}^2 + \|h_1\|_{H^{-1/2}}^2}\sqrt{\|f_2\|_{H^{1/2}}^2 +\|h_2\|_{H^{-1/2}}^2}\\
&=\|F_1\|_{H^{1/2}\oplus H^{-1/2}}\|F_2\|_{H^{1/2}\oplus H^{-1/2}}
\end{flalign*}
\end{proof}The sympletic form defined here is time invariant.
\begin{lemma}
For $F_t=U(t,1)F$, $\sigma(F_{1t},F_{2t}) = \sigma(F_1,F_2)$
\end{lemma}
\begin{proof}
Given $F_1, F_2 \in H^{1/2}(\mathbb{R}^3) \oplus H^{-1/2}(\mathbb{R}^3)$, there exists $\psi_1,\psi_2 \in L^2(\mathbb{R}^3)$ so that $F_1 =K^{-1}\psi_1, F_2 =K^{-1}\psi_2$.  Since $\mathcal{S}(\mathbb{R}^3)$ is a dense subspace of $L^2(\mathbb{R}^3)$, there are $\psi_{1j}\in \mathcal{S}(\mathbb{R}^3)\rightarrow\psi_1,\psi_{2j}\in \mathcal{S}(\mathbb{R}^3)\rightarrow\psi_2$.  Then $F_{1j} =K^{-1}\psi_{1j}\rightarrow F_1$ and $F_{2j} =K^{-1}\psi_{2j}\rightarrow F_2$ are convergent sequences, one has by the continuity of $\sigma$
\begin{equation*}
\sigma(F_1,F_2)=\lim_{\substack{j\rightarrow \infty}}\sigma(F_{1j},F_{2j})
\end{equation*}
Also the convergence $(F_{1j})_t\rightarrow(F_{1})_t$ follows from the continuity of $U(t,1)$  As a result,
\begin{equation*}
\sigma((F_1)_t,(F_2)_t)=\lim_{\substack{j\rightarrow \infty}}\sigma((F_{1j})_t,(F_{2j})_t)
\end{equation*} 
Since the test functions $F_{jt}$ are smooth, one can apply Green's identity to obtain
\begin{equation*}
\sigma((F_{1j})_t,(F_{2j})_t)=\sigma(F_{1j},F_{2j})
\end{equation*}
Let $j\rightarrow\infty$ to get the result.
\end{proof}
\section{Properties of Quantum Field}
\subsection{Definition of the Fields}
Our goal in this section is define a quantum field with dynamics of Klein-Gordon equation and satisfying the canonical communtation realtions(CCR) at $t=1$.  We work in a Hilbert space corresponding to "Euclidean" vacuum and develop properties of field such as invariance of correlation function under the isometry group for de-Sitter space.  

The Klein-Gordon field is too singular as a function of spacetime coordinates. The field instead is taken as an operator valued distribution, in other words a function of real test functions, $f,h\in\mathcal{S}(\mathbb{R}^3)$. The CCR have the form
\begin{equation*}
[\phi(h), \pi(f)] = i(h,f)
\end{equation*}
Or if the field is expressed as a sympletic form we have for $F=(f,h)$
\begin{equation*}
\sigma(\Phi, F) = \phi(h) - \pi(f)
\end{equation*} 
Then the CCR have the form
\begin{equation*}
[\sigma(\Phi, F_1), \sigma(\Phi, F_2)] = i\sigma(F_1, F_2)
\end{equation*}
Then taking into account the invariance of the sympletic form under time evolution, the time evolved field is defined as
\begin{flalign*}
\sigma(\Phi_t, F) & = \sigma(U_0(t,1)\Phi, F)\\
& = \sigma(\Phi, U_0(1,t)F)
\end{flalign*}

Next we pick a particular representation of CCR which is equivalent to choosing a specific vacuum. Let $\mathcal{H}$ be a Hilbert space and for $h\in\mathcal{H}$ let $a^*(h)$ and $a(h)$ be creation and annihilation operators on the Fock space $\mathcal{F}(\mathcal{H})$.  They are respectively linear and anti-linear in $h$ and satisfy
\begin{equation*}
[a(h_1), a^*(h_2)]=(h_1,h_2)
\end{equation*}
With $\mathcal{H} = L^2(\mathbb{R}^3, d\vec{p})$ we define
\begin{equation}
\sigma(\Phi, F) = i[a(K'F)-a^*(K'F)]
\end{equation}
where the renormalized $K'=\frac{2R}{\sqrt{\pi}}K$.  This field will be shown to satisfy the CCR after establishing a lemma. For $F_1, F_2\in H^{1/2}(\mathbb{R}^3) \oplus H^{-1/2}(\mathbb{R}^3)$, the sympletic form $\sigma(F_1, F_2)$ is well defined as shown in lemma 4.  For $\psi_1,\psi_2\in L^2(\mathbb{R}^3)$, the sympletic form on $L^2(\mathbb{R}^3)$ is $2\text{Im}(\psi_1,\psi_2)$.
\begin{lemma}
$K'$ is sympletic from $H^{1/2}(\mathbb{R}^3) \times H^{-1/2}(\mathbb{R}^3)$ to $L^2(\mathbb{R}^3)$.
\end{lemma}
\begin{proof}
One has from (14)
\begin{flalign*}
K'F & =\frac{2R}{\sqrt{\pi}W(1)}det \begin{pmatrix} \tilde{f} & \bar{v}(1,k)\\ R\tilde{h} & \bar{v}'(1,k) \end{pmatrix}\\
& = \frac{2R}{\sqrt{\pi}W(1)}[\tilde{f}\bar{v}'(1,k)-R\tilde{h}\bar{v}(1,k)]
\end{flalign*}
Since $f$ is real, $\overline{\tilde{f}(\vec{k})} = \tilde{f}(-\vec{k})$.  Then we have for $F_1=(f_1,h_1)$, $F_2=(f_2,h_2)$
\begin{flalign*}
2i\text{Im}(K'F_1, K'F_2) & = (K'F_1, K'F_2)-\overline{(K'F_1, K'F_2)}\\
& = \frac{4R^2}{\pi|W(1)|^2}\int[\tilde{f}_1(-\vec{k})\tilde{f}_2(\vec{k})v'\bar{v}' + R^2\tilde{h}_1(-\vec{k})\tilde{h}_2(\vec{k})v\bar{v}\\
&\quad\quad\quad\quad\quad\quad-\tilde{f}_1(\vec{k})\tilde{f}_2(-\vec{k})v'\bar{v}' - R^2\tilde{h}_1(\vec{k})\tilde{h}_2(-\vec{k})v\bar{v}\\
&\quad\quad\quad\quad\quad\quad+R\tilde{f}_1(\vec{k})\tilde{h}_2(-\vec{k})v\bar{v}'+R\tilde{h}_1(\vec{k})\tilde{f}_2(-\vec{k})v'\bar{v}\\
&\quad\quad\quad\quad\quad\quad-R\tilde{f}_1(-\vec{k})\tilde{h}_2(\vec{k})v'\bar{v}-R\tilde{h}_1(-\vec{k})\tilde{f}_2(\vec{k})v\bar{v}']\mathrm{d}\vec{k}
\end{flalign*}
The first four terms vanish after a change of variable $\vec{k}\rightarrow -\vec{k}$ in the second line.  So
\begin{flalign*}
2i\text{Im}(K'F_1, K'F_2) & =\frac{4R^3}{\pi W(1)}\int[\tilde{f}_1(-\vec{k})\tilde{h}_2(\vec{k})-\tilde{h}_1(-\vec{k})\tilde{f}_2(\vec{k})]\mathrm{d}\vec{k}\\
& = R^3i\int[f_1(\vec{x})h_2(\vec{x})-h_1(\vec{x})f_2(\vec{x})]\mathrm{d}\vec{x}\\
& = i\sqrt{\text{det}(g^{(3)}(1))}\int[f_1(\vec{x})h_2(\vec{x})-h_1(\vec{x})f_2(\vec{x})]\mathrm{d}\vec{x}\\
& = i\sigma(F_1, F_2)
\end{flalign*}
\end{proof}
Now, we can show the CCR for the $\sigma(\Phi, F)$ defined in (29) as follows
\begin{flalign*}
[\sigma(\Phi, F_1), \sigma(\Phi, F_2)] & = [a(K'F_1), a^*(K'F_2)]-[a(K'F_2), a^*(K'F_1)]\\
& = 2i\text{Im}(K'F_1, K'F_2)\\
& = i\sigma(F_1, F_2)
\end{flalign*} 
where $K'$ sympletic is used from lemma 6.
\subsection{Invariance of Correlation functions under O(1,4) symmetry}
Taking $F=(0,f)$, the smeared field can be expressed as 
\begin{flalign*}
\phi(t, f) & = \sigma(\Phi_t, (0,f))\\
& = \sigma(\Phi, U_0(1,t)(0,f))\\  
& = i[a(K'U_0(1,t)(0,f)) - a^*(K'U_0(1,t)(0,f))]\\
& = i[a(K_t'(0,f)) - a^*(K_t'(0,f))]
\end{flalign*}
where $K_t'=\frac{2R}{\sqrt{\pi}}K_t$ and $K_t$ is given by (14), so that
\begin{equation}
\begin{split}
[K_t'(0,f)](\vec{k}) & = \frac{2R}{\sqrt{\pi}tW(t,k)}\text{det } \begin{pmatrix} 0 & \bar{v}(t,k)\\ R\tilde{f}(\vec{k}) & t\bar{v}'(t,k) \end{pmatrix}\\
& = -\frac{2R^2}{\sqrt{\pi}t}\frac{\bar{v}(t,k)}{W(t,k)}\tilde{f}(\vec{k})\\
& = \frac{\sqrt{\pi}}{2i}R^2t^{-3/2}\mathcal{H}_\nu^{(2)}(kt)\tilde{f}(\vec{k})
\end{split}
\end{equation}
where (12) was used for $W(t,k)$.  Constructing the two point function over a vaccum state $\Omega$,
\begin{equation}
\begin{split}
\left(\Omega, \phi(t,f)\phi(t',f')\Omega\right) & = \left(\Omega, \sigma(\Phi_t, (0,f))\sigma(\Phi_{t'}, (0,f'))\Omega\right)\\
& = (\Omega, [a^*(K_t'(0,f)) - a(K_t'(0,f))][a(K_{t'}'(0,f')) - a^*(K_{t'}'(0,f'))]\Omega)\\
& = \left(\Omega, [a(K_t'(0,f)) , a^*(K_{t'}'(0,f'))]\Omega\right)\\
& = i\left(K_t'(0,f), K_{t'}'(0,f')\right)\\
& = i \int \overline{K_t'(0,f)}K_{t'}'(0,f')\mathrm{d}\vec{k}\\
& = i\frac{\pi}{4}R^4(tt')^{-3/2}\int \mathcal{H}_\nu^{(1)}(kt)\mathcal{H}_\nu^{(2)}(kt')\tilde{f}(-\vec{k})\tilde{f}'(\vec{k})\mathrm{d}\vec{k}
\end{split}
\end{equation}
where (30) was used in the last line.  Now we want to find the pointwise two-point function defined as
\begin{equation*}
\mathcal{W}(t,\vec{x};t',\vec{y}) = (\Omega, \phi(t,\vec{x})\phi(t',\vec{y})\Omega)
\end{equation*} 
The pointwise two-point function can be turned into distribution through the test function $f(x)$ and the $3\times3$ spatial part of the de-Sitter metric, $\sqrt{\text{det}(g^{(3)}(t))}$ such that
\begin{equation*}
\phi(t,f) = \int_{\mathbb{R}^3}\phi(t,\vec{x})f(\vec{x})\sqrt{det(g^{(3)}(t))}\mathrm{d}\vec{x}
\end{equation*}
Then the two point function can be written in terms of $\mathcal{W}(t,\vec{x};t',\vec{y})$,
\begin{equation}
\begin{split}
\left(\Omega, \phi(t,f)\phi(t',f')\Omega\right) & = \int f(\vec{x}) \sqrt{det(g^{(3)}(t))}\mathcal{W}(t,\vec{x};t',\vec{y})\sqrt{det(g^{(3)}(t'))}f'(\vec{y})\mathrm{d}\vec{x}\mathrm{d}\vec{y}\\
& = \left(\frac{R}{t}\right)^3\left(\frac{R}{t'}\right)^3\int f(\vec{x}) \mathcal{W}(t,\vec{x};t',\vec{y})f'(\vec{y})\mathrm{d}\vec{x}\mathrm{d}\vec{y}\\
\end{split}
\end{equation}
From (31) and (32), one can identify
\begin{equation*}
\mathcal{W}(t,\vec{x};t',\vec{y}) = i\frac{\pi}{4R^2}\frac{(tt')^{3/2}}{(2\pi)^3}\int e^{i\vec{k}\cdot(\vec{x}-\vec{y})} \mathcal{H}_\nu^{(1)}(kt)\mathcal{H}_\nu^{(2)}(kt')\mathrm{d}\vec{k}
\end{equation*}
which agrees with Schomblond-Spindel\cite{Schomblond}.  According to Schomblond-Spindel, this two-point function can be written in terms of $\Delta(x, y)$ and $\Delta^1(x, y)$
\begin{flalign*}
i\Delta(x,y) & = (\Omega,[\phi(x),\phi(y)]\Omega)\\
& = -\frac{i}{8\pi R^2}\frac{\frac{1}{4} - \nu^2}{\cos(\nu\pi)}\epsilon(t-t')\text{Im F}\left(-\nu + \frac{3}{2},\nu+\frac{3}{2};2;\frac{1+p}{2}\right)
\end{flalign*}and
\begin{flalign*}
\Delta^1(x,y) & = (\Omega,\{\phi(x),\phi(y)\}\Omega)\\
& = \frac{1}{8\pi R^2}\frac{\frac{1}{4} - \nu^2}{\cos(\nu\pi)}\text{Re F}\left(-\nu + \frac{3}{2},\nu+\frac{3}{2};2;\frac{1+p}{2}\right)
\end{flalign*}
where $p$ is defined as follows
\begin{flalign*}
p&=\frac{x\cdot y}{R^2}\\
&= \cosh \left(\frac{s}{R}\right)
\end{flalign*}
where $s$ is a geodesic distance $s=s(x,y)$ between two points in de-Sitter space.
Then
\begin{flalign*}
\mathcal{W}(x,y) & = \frac{1}{2}[i\Delta(x,y) +\Delta^1(x,y)]\\
& = \frac{1}{8\pi R^2}\frac{\frac{1}{4} - \nu^2}{\cos(\nu\pi)}\bigg[\text{Re F}\left(-\nu + \frac{3}{2},\nu+\frac{3}{2};2;\frac{1+p}{2}\right)\\&\quad-i\epsilon(t-t')\text{Im F}\left(-\nu + \frac{3}{2},\nu+\frac{3}{2};2;\frac{1+p}{2}\right)\bigg]
\end{flalign*}
Thus the two-point function depends only on the geodesic distance. Thus it is invariant under the isometry group $O(1,4)$ for de-Sitter space.  All correlation functions can be expressed in terms of the two point function so they are invariant too.
\subsection{Locality Condition}
We want to show our field has a local commutator.

First we recall some definitions on any time oriented Lorentzian manifold $(M, g)$.  Let's define the causal future and past of a point $p$ in the manifold by
\begin{equation*}
J^\pm(p) = \{q\in M:\exists \text{ a future/past directed causal curve from } p \text{ to } q\}
\end{equation*}
and
\begin{definition}
A Lorentzian manifold $(M,g)$ is globally hyperbolic if the following equivalent conditions hold\\
1.  For any $p,q \in M$ the set $J^+(p) \cap J^-(q)$ is compact.\\
2.  There is Cauchy surface which is a spacelike hypersurface intersected by every causal curve exactly once.\\
3.  $(M,g)$ is diffeomorphic with a manifold $(\mathbb{R}\times\Sigma,g')$ for which $\Sigma_t=\{t\}\times\Sigma$ is a Cauchy surface for all $t$.
\end{definition}
We are now in a position to quote a theorem\cite{BGP07}
\begin{theorem}
Let $(M,g)$ be globally hyperbolic and let $(\Sigma, n)$ be a Cauchy surface with normal $n$.  Then for any $f,h\in C_0^\infty(\Sigma)$ there exists a unique regular solution $u\in C^\infty(\mathbb{R}\times\Sigma)$ of $(-\Box +m^2)u=0$ such that $u_\Sigma=f$ and $\nabla_nu|_\Sigma=h$.  

Moreover, $supp(u)\subset J^\pm(supp(f)\cup supp(h))$.
\end{theorem}
De-Sitter space is conformally equivalent to the subspace $(0,\infty)\times \mathbb{R}^3$ of Minkowski space as seen in (2). Hence, $J^\pm(p)$ is the same for de-Sitter space as for the subspace ofMinkowski space.  Therefore de-Sitter space is globally hyperbolic just like the subspace of Minkowski space.  The compact condition of $J^+(p) \cap J^-(q)$ is easy to check.  Also our solution is $C^{\infty}$ by lemma 2 so it must be the one refered to in the theorem.  Therefore our solution has the claimed support properties. Now we can demonstrate the locality condition.
\begin{theorem}
If $J^\pm(\{t_1\}\times supp(F_1)) \cap \{t_2\} \times supp(F_2) = \emptyset$,
\begin{equation*}
[\sigma(\Phi_{t_1}, F_1), \sigma(\Phi_{t_2}, F_2)] = 0\\
\end{equation*}
\end{theorem}
\begin{proof}
We compute
\begin{flalign*}
[\sigma(\Phi_{t_1}, F_1), \sigma(\Phi_{t_2}, F_2)] & = i\sigma(U(1,t_1)F_1, U(1, t_2)F_2)\\
& = i \sigma(U(t_2, t_1)F_1, F_2)\\
& = 0
\end{flalign*}
since $supp(U(t_2, t_1)F_1) \cap supp(F_2) = \emptyset$. To see $J^\pm(\{t_1\}\times supp(F_1)) \cap \{t_2\} \times supp(F_2) = \emptyset$ implies $supp(U(t_2, t_1)F_1) \cap supp(F_2) = \emptyset$, consider the $t_2$ slice of $J^\pm(\{t_1\}\times supp(F_1))$ which is equal to $\{t_2\}\times (supp(F_1))_{t_2-t_1}$ where $X_t=\{y\in\mathbb{R}^3:d(y,X)\leq t\}$.  That implies,
\begin{equation*}
(supp(F_1))_{t_2-t_1} \cap supp(F_2) = \emptyset
\end{equation*}
Since $supp(U(t_2,t_1)F_1)\subset(supp(F_1))_{t_2-t_1}$ by theorem 2,
\begin{equation*}
supp(U(t_2, t_1)F_1) \cap supp(F_2) = \emptyset
\end{equation*}
\end{proof}
This commutation relation states the causality relation between the fields which commute if $F_1$ and $F_2$ have spacelike separated supports.
\section{Unitary Implementability of Time Evolution Operators}
A Segal field is the operator on the Fock space $\mathcal{F}(\mathcal{H})$ by
\begin{equation*}
\Phi_s(\psi) = i[a(\psi)-a^*(\psi)],\quad \psi\in\mathcal{H}
\end{equation*}
It is real linear in $\mathcal{H}$ and satisfies the commutation relation,
\begin{equation*}
[\Phi_s(\psi_1),\Phi_s(\psi_2)] = 2i\text{Im}(\psi_1,\psi_2)
\end{equation*}  
If an operator, $T:\mathcal{H}\rightarrow \mathcal{H}$ is real linear and sympletic such that $\text{Im}(T\psi_1, T\psi_2) = \text{Im}(\psi_1, \psi_2)$, then $\Phi_s'(\psi) = \Phi_s(T\psi)$ also satisfies the commutation relation.  The question of unitary implementablility of this transformation is answered by Shale's theorem\cite{Reed, Shale}.
\begin{theorem}
Let $T$ be invertible sympletic transformation.  Then there exists $\mathcal{U}_T$ on $\mathcal{F}(\mathcal{H})$ so that $\Phi'_s(\psi) = \mathcal{U}_T\Phi_s(\psi)\mathcal{U}_T^{-1}$, iff $T^\dagger T-I$ is Hilbert-Schmidt.
\end{theorem}
Here $T^\dagger$ is the real adjoint of $T$ such that $\text{Re}(\psi_1, T\psi_2)=\text{Re}(T^\dagger\psi_1, \psi_2)$. 

We are interested in the case, $\mathcal{H} = L^2(\mathbb{R}^3)$, the field
\begin{equation*}
\sigma(\Phi, F) = \Phi_s(K'F)
\end{equation*}
and the time evolution with no potential
\begin{equation*}
\sigma(\Phi_t, F) =\sigma(\Phi, U_0(1,t)F)=\Phi_s(K'U_0(1,t)F)
\end{equation*}
\begin{theorem}
There is no interval $I\subset(0,\infty)$ containing $1$ so that for $t\in I$ there exists a unitary operator, $\mathcal{U}_t \in \mathcal{F}(\mathcal{H})$ satisfying, $\mathcal{U}_t^{-1}\sigma(\Phi, F)\mathcal{U}_t=\sigma(\Phi, U_0(1,t)F)$.
\end{theorem}
\begin{proof}Let's determine the operator $T$ stated in theorem 2 as follows.  We would like
\begin{equation*}
\mathcal{U}_t^{-1}\Phi_s(K'F)\mathcal{U}_t=\Phi_s(K'U_0(1,t)F)
\end{equation*}
Let $G=K'U_0(1,t)F$, or $F= U_0(t,1)K'^{-1}G$, then this becomes
\begin{equation*}
\mathcal{U}_t^{-1}\Phi_s(K'U_0(t,1)K'^{-1}G)\mathcal{U}_t=\Phi_s(G)
\end{equation*}
Replacing $K'$ by $K$ this is the same as
\begin{equation*}
\Phi_s(\hat{U}_0(t,1)G) = \mathcal{U}_t\Phi_s(G)\mathcal{U}_t^{-1}
\end{equation*}
so $T=\hat{U}_0(t,1)$ is the time evolution operator we saw in section 3. 

Let $Z = \hat{U}^\dagger_0(t,1)\hat{U}_0(t,1) - I$ and $\hat{U}^\dagger_0(t,1):L^2(\mathbb{R}^3)\rightarrow L^2(\mathbb{R}^3)$.  For $\chi, \psi \in L^2(\mathbb{R}^3)$,
\begin{flalign*}
\text{Re}(\hat{U}^\dagger_0(t,1)\chi, \psi) & = \text{Re}(\chi, \hat{U}_0(t,1)\psi)\\
& = \text{Re}(\chi, KU(t,1)K^{-1}\psi)\\
& = -\text{Im}(KK^{-1}i\chi, KU(t,1)K^{-1}\psi)\\
& = -\frac{1}{2}\sigma(K^{-1}i\chi, U(t,1)K^{-1}\psi)\\
& = -\frac{1}{2}\sigma(U(1, t)K^{-1}i\chi, K^{-1}\psi)\\
& = -\text{Im}(KU(1,t)K^{-1}i\chi, \psi)\\
& = -\text{Re}(-i(KU(1,t)K^{-1}i\chi, \psi))\\
& = -\text{Re}(iKU(1,t)K^{-1}i\chi, \psi))
\end{flalign*}
where I have used the identity $U^{-1}(t,1)=U(1,t)$ shown at the end of section 3.  Then one has
\begin{equation*}
\hat{U}^\dagger_0(t,1) = -i\hat{U}_0(1,t)i
\end{equation*}
Then
\begin{equation}
\begin{split}
Z & = -i\hat{U}_0(1,t)i\hat{U}_0(t,1) - I\\
& = -i\hat{U}_0(1,t)(i\hat{U}_0(t,1) - \hat{U}_0(t,1)i)
\end{split}
\end{equation}
Since $\hat{U}_0(t,1)$ is bounded with bounded inverse, $Z$ is Hilbert-Schmidt iff $i\hat{U}_0(t,1) - \hat{U}_0(t,1)i$ is Hilbert-Schmidt.  From (17) we have
\begin{equation}
(i\hat{U}_0(t,1) - \hat{U}_0(t,1)i)\psi(\vec{k}) = \frac{2i}{W(1,k)} \text{det} \begin{pmatrix} \bar{v}(t,k) & \bar{v}(1,k)\\ t\bar{v}'(t,k) & \bar{v}'(1,k) \end{pmatrix}\overline{\psi(-\vec{k})}
\end{equation}
Since $i\hat{U}_0(t,1) - \hat{U}_0(t,1)i$ is a multiplication operator, it is Hilbert-Schmidt iff it is zero.  

Suppose there is an interval $I$ so that $i\hat{U}_0(t,1) - \hat{U}_0(t,1)i=0$ for $t\in I$.  Then
\begin{equation*}
\bar{v}(t,k)\bar{v}'(1,k)-t\bar{v}'(t,k)\bar{v}(1,k)=0
\end{equation*}
Case 1, $v(1,k)=0$.  If $v'(1,k)=0$, then $v(t,k)$ has zero initial data and thus vanishes whish is false.  So $v'(1,k)\neq0$ and then the equation implies $v(t,k)=0$ which is again false.\\
Case 2, $v(1,k)\neq0$.  One has,
\begin{equation*}
\bar{v}'(t,k) = \frac{1}{t}\gamma\bar{v}(t,k)
\end{equation*}
where
\begin{equation*}
\gamma = \frac{\bar{v}'(1,k)}{\bar{v}(1,k)}
\end{equation*}
After integrating from 1 to $t$, one has
\begin{flalign*}
\text{ln}\left[\frac{\bar{v}(t,k)}{\bar{v}(1,k)}\right]&=\gamma\text{ln}(t)\\
\bar{v}(t,k)& = t^\gamma\bar{v}(1,k)
\end{flalign*}
which is false because $t^{3/2}\mathcal{H}_\nu(kt) \neq t^\gamma \mathcal{H}_\nu(k)$.

Thus $i\hat{U}_0(t,1) - \hat{U}_0(t,1)i\neq0$ and $Z$ is not Hilbert-Schmidt.  Then Shale's theorem proves theorem 5.
\end{proof}
Next we consider the analysis of unitary transformation for the time evolved field from no potential to a scalar potential.
\begin{theorem}
Given a $V\in C_0^\infty([0,t_0]\times\mathbb{R}^3)$, then for $t\geq t_0$ there exists a unitary operator, $\mathcal{U}_t \in \mathcal{F}(\mathcal{H})$, so that $\sigma(\Phi, U_V(1,t)F)=\mathcal{U}_t^{-1}\sigma(\Phi, U_0(1,t)F)\mathcal{U}_t$.
\end{theorem}
\noindent Remarks: This proof is similar to the proof of Reed and Simon \cite{Reed} on Minkowski space.  We take times away from the perturbation for convenience and because it is the most interesting case.  However it is probably not necessary.
\begin{proof}
Let's determine the operator $T$ again as follow.  We would like
\begin{equation*}
\mathcal{U}_t^{-1}\Phi_s(K'U_0(1,t)F)\mathcal{U}_t=\Phi_s(K'U_V(1,t)F)
\end{equation*}
Let $G=K'U_V(1,t)F$, or $F= U_V(t,1)K'^{-1}G$, then this becomes
\begin{equation*}
\mathcal{U}_t^{-1}\Phi_s(K'U_0(1,t)U_V(t,1)K'^{-1}G)\mathcal{U}_t=\Phi_s(G)\\
\end{equation*}
So we want
\begin{equation*}
\Phi_s(TG) = \mathcal{U}_t\Phi_s(G)\mathcal{U}_t^{-1}
\end{equation*}
where
\begin{equation*}
T = K'U_0(1,t)U_V(t,1)K'^{-1} = KU_0(1,t)U_V(t,1)K^{-1}
\end{equation*}
Infering $T^\dagger = -iT^{-1}i$ as in the proof of theorem 3, then $Z'=T^\dagger T - I = -iT^{-1}(iT-Ti)$.  By our earlier results $T$ is bounded so $Z'$ is Hilbert-Schmidt if $iT - Ti$ is Hilbert-Schmidt.  Given the covergent series in (24) for $U_V(t,1)$, we get a convergent series for $T$ which states
\begin{flalign*}
T & = I- \int_{1}^{t}KU_0 (1,s)V(s)U_0(s,1)K^{-1}\mathrm{d}s +\cdots\\
& = I - \int_{1}^{t}K_sV(s)K_s^{-1}\mathrm{d}s +\cdots\\
& = I - \int_{1}^{t}\hat{V}(s)\mathrm{d}s +\cdots
\end{flalign*}
In analogy to (27) with $K$ replaced by $K_s$ from (14), and $\psi \in L^2(\mathbb{R}^3)$, 
\begin{equation}
\begin{split}
\hat{V}(s)\psi(\vec{p}) & = K_sV(s)K_s^{-1}\psi(\vec{p})\\
& = -\frac{(2\pi)^{3/2}R}{s}\frac{\bar{v}(s, p)}{W(s,p)}(\tilde{V}(s,\cdot)\ast\tilde{f}_s)(\vec{p})\\
& = -\frac{(2\pi)^{3/2}R}{s}\frac{\bar{v}(s, p)}{W(s,p)}\int\tilde{V}(s,\vec{p}-\vec{q})[v(s,q)\psi(\vec{q}) + \bar{v}(s,q)\overline{\psi(-\vec{q})}]\mathrm{d}\vec{q}
\end{split}
\end{equation} 
As shown in the proof of theorem 1, $\hat{V}(s)$ is bounded on $L^2(\mathbb{R}^3)$.  

More generally,
\begin{equation*}
T = \overset{\infty}{\sum_{\substack{n=0}}} T_n
\end{equation*}
where with $T_0=I$ and
\begin{equation}
T_n = (-1)^n\int_{1}^{t}\cdots\int_{1}^{s_{n-1}}\hat{V}(s_1)\cdots\hat{V}(s_n)\mathrm{d}s_n\cdots\mathrm{d}s_1
\end{equation}

Next, let's look at the Hilbert-Schmidt condition of $iT - Ti$ for the first two terms.  In accordance with (35) and (36),
\begin{flalign*}
\overset{1}{\sum_{\substack{n=0}}}(iT_n -T_ni)\psi(\vec{p}) & = (iT_1-T_1i)\psi(\vec{p})\\ &=2(2\pi)^{3/2}iR\int\tau(\vec{p},\vec{q})\overline{\psi(\vec{q})}\mathrm{d}\vec{q}
\end{flalign*}
where the kernel is
\begin{equation*}
\tau(\vec{p},\vec{q})=\int\frac{\bar{v}(s, p)}{sW(s,p)}\tilde{V}(s,\vec{p}+\vec{q})\bar{v}(s,q)\mathrm{d}s
\end{equation*}
Since $W(s,p)$ is independent of $p$ and finitely dependent on $s$ as seen from (12), we choose the simplifed kernel to be
\begin{equation*}
\tau'(\vec{p},\vec{q})=\int_1^t\bar{v}(s, p)\tilde{V}(s,\vec{p}+\vec{q})\bar{v}(s,q)\mathrm{d}s
\end{equation*}
Then $iT_1-T_1i$ is Hilbert-Schmidt or not depends on the condition,
\begin{equation}
\iint|\tau'(\vec{p},\vec{q})|^2\mathrm{d}\vec{q}\mathrm{d}\vec{p} < \infty
\end{equation}
Now we analyze the kernel in three different regions of momentum, with $p=|\vec{p}|$ and $q=|\vec{q}|$\\
(1) $p$ and $q$ $\leq 1$,  one has, from lemma 1, that $|\bar{v}(s, q)|$, $|\bar{v}(s, p)|$ are bounded.  Also
$|\tilde{V}(\vec{p}+\vec{q},s)|$ is bounded.  Hence $\tau'(\vec{p},\vec{q})$ is bounded as well.  So
\begin{equation*}
\iint_{p,q\leq1}|\tau'(\vec{p},\vec{q})|^2\mathrm{d}\vec{p}\mathrm{d}\vec{q} < \infty
\end{equation*}\\
(2) $p$ and $q\geq 1$, one has the asymptotic expression of Hankel function,
\begin{flalign*}
\bar{v}(s, p) & = s^{3/2}e^{\mu\pi/2}H_\nu^{(2)}(ps)\\
& = \sqrt{\frac{2}{\pi p}}se^{-i(ps - \frac{\pi}{4})}\left[1 - \frac{i(4\nu^2 -1)}{8ps} + \cdots\right]
\end{flalign*}
Now the kernel, up to a multiplicative constant, is split into six terms such that $\tau'= \tau_0' +\tau_1' + \cdots + \tau_5'$ given by
\begin{multline*}
\tau'(\vec{p}+\vec{q}) = \frac{1}{\sqrt{pq}}\int_1^t e^{-i(p+q)s}s^2\tilde{V}(s,\vec{p}+\vec{q})\bigg[1 - \frac{a}{ps}- \frac{a}{qs} + \frac{a^2}{pqs^2} \\+ \mathcal{O}\left(\frac{1}{(ps)^2}\right) + \mathcal{O}\left(\frac{1}{(qs)^2}\right)\bigg]\mathrm{d}s
\end{multline*}
where $a = \frac{i}{8}(4\nu^2 -1)$.  Consider the first term of the series
\begin{flalign*}
\tau_0'(\vec{p}+\vec{q}) & = \frac{1}{\sqrt{p_0q_0}}\int_1^t e^{-i(p_0+q_0)s}s^2\tilde{V}(s,\vec{p}+\vec{q})\mathrm{d}s\bigg|_{\substack{p_0=p\\q_0=q}}\\
&=-\frac{1}{\sqrt{p_0q_0}}\frac{\partial^2}{\partial p_0^2}\int_1^t e^{-i(p_0+q_0)s}\tilde{V}(s,\vec{p}+\vec{q})\mathrm{d}s\bigg|_{\substack{p_0=p\\q_0=q}}\\
&=-\frac{1}{\sqrt{p_0q_0}}\frac{\partial^2\tilde{V}}{\partial p_0^2}(p_0+q_0,\vec{p}+\vec{q})\bigg|_{\substack{p_0=p\\q_0=q}}
\end{flalign*}
Since the full Fourier transform $\tilde{V}$ is Schwarz function, so is its second derivative, and it is bounded by $(|\vec{p}+\vec{q}|+p+q)^{-N}$ for any $N$.  Then we have
\begin{flalign*}
|\tau_0'(\vec{p}+\vec{q})| & \leq \frac{C_t}{\sqrt{pq}}(|\vec{p}+\vec{q}|+p+q)^{-N}\\
& \leq \frac{C_t}{\sqrt{pq}}(p+q)^{-N}\\
& \leq  \frac{C_t}{p^{(N+1)/2}q^{(N+1)/2}}
\end{flalign*}
This gives a finite result when square-integrated over $p$, $q \geq 1$, provided $N>2$. The same analysis holds for the next three terms $\tau_1'$, $\tau_2'$ and $\tau_3'$.  For the fifth term,
\begin{flalign*}
|\tau_4'(\vec{p}+\vec{q})| & \leq \frac{1}{\sqrt{pq}}\int_1^t |\tilde{V}(s,\vec{p}+\vec{q})|\mathcal{O}\left(\frac{1}{p^2}\right)\mathrm{d}s\\
& \leq \frac{C_t}{\sqrt{pq}}\underset{s}{\sup}|\tilde{V}(s,\vec{p}+\vec{q})|\mathcal{O}(p^{-2})
\end{flalign*}
But for any $N$ there is a constant $C$ so
\begin{equation}
|\tilde{V}(s,\vec{p}+\vec{q})|\leq C(|\vec{p}+\vec{q}|+1)^{-N}
\end{equation}
Now the kernel integrated over $\vec{q}$ will be
\begin{flalign*}
\int|\tau_4'(\vec{p}+\vec{q})|^2\mathrm{d}\vec{q} & \leq C_t^2\mathcal{O}(p^{-5})\int(1+|\vec{p}+\vec{q}|)^{-2N}\mathrm{d}\vec{q}\\
& = C_t'\mathcal{O}(p^{-5})\quad \text{for } N>\frac{3}{2}
\end{flalign*}
Then the integral over $\vec{p}$ is finite since $\mathcal{O}(p^{-5})$ decays fast enough to be convergent.  The same result is obtained for the last term, $\tau_5'$ with $p$ and $q$ interchanged.  Therefore, (37) holds.\\
(3) $q\leq 1$ and $p\geq 1$, one has, from lemma 1,
\begin{equation*}
|\bar{v}(s, p)|\leq C s^{3/2}\omega^{-1/2}(p)
\end{equation*}
Combined with (38) this gives
\begin{equation*}
|\tau'(\vec{p}+\vec{q})| \leq C_t(1+|\vec{p}+\vec{q}|)^{-N}\omega^{-1/2}(q)\omega^{-1/2}(p)
\end{equation*}
Then
\begin{equation*}
\int_{|\vec{p}|\geq 1}|\tau'(\vec{p}+\vec{q})|^2\mathrm{d}\vec{p}\leq \frac{C_t^2}{\sqrt{2}}\omega^{-1}(q)\int(1+|\vec{p}+\vec{q}|)^{-2N}\mathrm{d}\vec{p}
\end{equation*}
This is finite, provided $N>\frac{3}{2}$. Then the integral over $|\vec{q}|\leq1$ is finite as well.  So (37) holds again.\\ 
(4)$p\leq 1$ and $q\geq 1$. The same analysis as (3) is used.  Hence, $iT_n-T_ni$ is Hilbert-Schmidt for the $n=0,1$ terms.\\ 

In general, the Hilbert-Schmidt norm of $iT-Ti$ satisfies
\begin{equation}
\|iT-Ti\|_{2} \leq \|iT_1-T_1i\|_{2}+2\overset{\infty}{\sum_{\substack{n=2}}}\|T_n\|_{2}
\end{equation}
where $T_n$ is given by (36).  Then we have for $n\geq2$
\begin{flalign*}
\|T_n\|_{2} & \leq \int_{1}^{t}\cdots\int_{1}^{s_{n-1}}\|\hat{V}(s_1)\|_2\cdots\|\hat{V}(s_n)\|_2\mathrm{d}s_n\cdots\mathrm{d}s_1\\
&\leq\frac{M_2^n(t-1)^n}{n!}
\end{flalign*}
where $M_2 \equiv \text{sup}\|\hat{V}(s)\|_2$.  However we cannot show $M_2$ is finite.  We might try it as follows.  According to (35) and lemma 1, $\hat{V}(s)$ has a kernel, 
\begin{equation}
|\hat{V}(\vec{p} - \vec{q},s)| \leq C\omega^{-1/2}(\vec{p})(\tilde{V}(s,\vec{p} - \vec{q})| +|\tilde{V}(s,\vec{p} + \vec{q})|)\omega^{-1/2}(\vec{q})
\end{equation}
Hence, we have
\begin{equation*}
\|\hat{V}(s)\|_2^2 \leq C\iint|\tilde{V}(s,\vec{p} - \vec{q})|^2 (1+p)^{-1}(1+q)^{-1}\mathrm{d}\vec{p}\mathrm{d}\vec{q}
\end{equation*}
which is infinite since it does not have sufficient decay in $\vec{p}+\vec{q}$.

Instead, let's rewrite the Hilbert-Schmidt norm for $n\geq2$ as
\begin{flalign*}
\|T_n\|_{2} & \leq \int_{1}^{t}\cdots\int_{1}^{s_{n-1}}\|\hat{V}(s_1)\hat{V}(s_2)\|_2\cdots\|\hat{V}(s_{n-1})\hat{V}(s_n)\|_2\mathrm{d}s_n\cdots\mathrm{d}s_1\\
&\leq\int_{1}^{t}\cdots\int_{1}^{s_{n-1}}\|\hat{V}(s_1)\|_4\|\hat{V}(s_2)\|_4\cdots\|\hat{V}(s_{n-1})\|_4\|\hat{V}(s_n)\|_4\mathrm{d}s_n\cdots\mathrm{d}s_1\\
&\leq\frac{M_4^n(t-1)^n}{n!}
\end{flalign*}
where $M_4 \equiv \text{sup}\|\hat{V}(s)\|_{\mathcal{I}_4}$ is shown to be finite below. This expression holds for both even $n$ and odd $n$ since $\|\hat{V}(s_n)\|\leq\|\hat{V}(s_n)\|_4$ for odd $n$.  To see $M_4$ finite, we take (38) again and compute the $\mathcal{I}_4$ norm that is defined as $\|\hat{V}(s)\|_{\mathcal{I}_4} = [\text{Tr}(|\hat{V}|^4)]^{1/4}$.
\begin{flalign*}
\|\hat{V}\|_4^4 & = \text{Tr}(|\hat{V}|^4)\\
& = \text{Tr}(\hat{V}^*\hat{V}\hat{V}^*\hat{V})\\
& = \bigg|\int\hat{V}^*(p_1, p_2)\hat{V}(p_2, p_3)\hat{V}^*(p_3, p_4)\hat{V}(p_4, p_1)\mathrm{d}\vec{p_1}\cdots\mathrm{d}\vec{p_4}\bigg|\\
& \leq C \int(1+p_1)^{-1}|\tilde{V}(s,\vec{p_1} - \vec{p_2})|\cdots (1+p_4)^{-1}|\tilde{V}(s,\vec{p_4} - \vec{p_1})|\mathrm{d}\vec{p_1}\cdots\mathrm{d}\vec{p_4}
\end{flalign*}
Take the region $p_1<p_2,p_3,p_4$.  The integral over this region is less than
\begin{equation*}
\int(1+p_1)^{-4}|\tilde{V}(s,\vec{p_1} - \vec{p_2})||\tilde{V}(s,\vec{p_2} - \vec{p_3})||\tilde{V}(s,\vec{p_3} - \vec{p_4})||\tilde{V}(s,\vec{p_4} - \vec{p_1})|\mathrm{d}\vec{p_1}\cdots\mathrm{d}\vec{p_4}
\end{equation*}
Drop the last factor and do integrals over $\vec{p_4}$, $\vec{p_3}$,$\vec{p_2}$,$\vec{p_1}$ in that order.  Other regions like  $p_2<p_1,p_3,p_4$ are similar.  So $M_4$ is finite.  Now,
\begin{equation*}
\overset{\infty}{\sum_{\substack{n=2}}}\|T_n\|_{2} \leq \overset{\infty}{\sum_{\substack{n=2}}}\frac{M_4^n(t-1)^n}{n!} < e^{M_4(t-1)} < \infty
\end{equation*}
Since $iT_1 - T_1 i$ and $T_n$ for $n\geq2$ are Hilbert-Schmidt, $iT-Ti$ is Hilbert-Schmidt, so $Z'$ is Hilbert-Schmidt. $T$ is unitarily implementable by theorem 2.
\end{proof}
\section{Conclusion}
The result of theorem 5 that time evolution is not unitarily implementable means there is no good particle concept.  In theorem 6, the relative time evolution is unitarily implementable in the sense that fields at a fixed time can be unitarily transformed from no potential to a scalar potential. This means that a local potential has a limited effect on the dynamics of the field.  This property may be useful in future investigations of field theory on de-Sitter space.

In this article, only local scalar potential has been treated.  Local vector potentials and gravitational perturbations are also interesting but are not pursued here.  For some related results in Minkowski space see Dimock\cite{Dimock01,Dimock02}.
\section{Acknowlegment}
I would like to take this opportunity to thank Dr. Dimock's dedicated effort of teaching and his valuable advice.


\begin{thebibliography}{15}
\bibitem{Chernikov}
N.A. Chernikov and E.A. Tagirov, Quantum theory of Scalar field in de Sitter space-time, Annales de l'Institut Henri Poincare, A Vol.9, Issue 2, p.109, 1968.
\bibitem{Nachtmann}
O. Nachtmann, Quantum Theory in de-Sitter space, Communication of Mathematical Physics., Vol.6, p.1, 1967.
\bibitem{Allen}
B. Allen, Physical Review D, Vol.32, No.12, p.3136, 1985.
\bibitem{Schomblond}
C. Schomblond and P. Spindel, Ann. Inst. Henri Poincare, Vol.A25, p.67, 1976.
\bibitem{Bros}
J. Bros and U. Moschella, Reviews of Math, Phys., Vol.8, No.3, p.327, 1996.
\bibitem{Kay}
B.S. Kay and R. Wald, Physics Reports, Vol. 207, p.49, 1991
\bibitem{Gazeau}
J. Gazeau and M. Rey, Fifth International Conference on Mathematical Methods in Physics (IC2006), Rio de Janeiro : Br\'esil (2006).
\bibitem{Watson}
G. Watson, \emph{A Treatise on the Theory of Bessel Functions}, Cambridge Univ. Press, 1966.
\bibitem{BGP07} C. B\"{a}r, N. Ginoux, F. Pf\"{a}ffle, {\em Wave
Equations on Lorentzian Manifolds and Quantization},
European Mathematical Society Publishing House, 2007.
\bibitem{Reed}
M. Reed and B. Simon, \emph{Methods of Modern Mathematical Physics III}, Academic Press, N.Y., 1979.
\bibitem{Dimock01}
J. Dimock, J. Math. Phys, Vol.20, p.1791, 1979.
\bibitem{Dimock02}
J. Dimock, J. Math. Phys, Vol.20, p.2549, 1979.
\bibitem{Shale}
D. Shale, Trans. Am. Math. Soc., Vol. 103, p.149, 1962.
\end{thebibliography}
\end{document}